\documentclass{ws-ijmpa}

\begin{document}

\markboth{A. Jevicki, K. Jin}
{Solitons and AdS String Solutions}

%
\catchline{}{}{}{}{}
%

\title{SOLITONS AND ADS STRING SOLUTIONS}

\author{ANTAL JEVICKI\footnote{
Speaker at the conference.}}

\address{Department of Physics, Brown University, Box 1843\\
Providence, RI 02912, USA\\
antal@het.brown.edu}

\author{KEWANG JIN}

\address{Department of Physics, Brown University, Box 1843\\
Providence, RI 02912, USA\\
jin@het.brown.edu}

\maketitle

\begin{history}
\received{Day Month Year}
\revised{Day Month Year}
\end{history}

\begin{abstract}
In this contribution we describe some soliton based techniques for generating classical AdS string solutions. The methods introduced are useful for further understanding of rotating AdS configurations with spikes which correspond to higher twist operators in SYM theory. The main identification (accomplished in arXiv:0712.1193) between solitons and string spikes is reviewed and extended. We describe how inverse scattering technique can be applied for reconstructing AdS string configurations from soliton solutions of sinh-Gordon theory (in the example of ${\rm AdS}_3$).

\keywords{Classical string solutions; Integrable systems; AdS/CFT.}

\end{abstract}

\ccode{PACS numbers: 11.25.Hf, 123.1K}

\section{Introduction}

For understanding the AdS/CFT correspondence at the string level it is useful to peruse construction of classical string configurations moving in AdS spacetime. The theory is classically (and probably also quantum mechanically) integrable and following the experience with two dimensional integrable field theories the construction of its exact solution might be possible. At present particular examples representing classical motions in AdS are known. The first example is given by the rotating folded string of GKP.\cite{GKP} It was established
to be dual to twist two (large) spin operators of Super Yang-Mills (SYM) theory whose anomalous dimensions are of interest for scattering processes of QCD. An extension of the folded string solution is given by the n-spike configuration constructed by Kruczenski.\cite{Kruczenski} These solutions are special in that they represent static configurations in a uniformly rotating reference frame. One 
therefore expects a potentially much larger set of solutions with nontrivial dynamics.

In this contribution we describe our recent effort\cite{JJKV} in developing techniques for generating AdS string solutions. As it has been much studied recently the string in AdS (or more generally in the full $AdS \times S$ spacetime) is integrable both at the classical and quantum level. In conformal type gauges the equations are that of nonlinear sigma models defined on the spaces in question. One well known approach to study integrability of sigma models and generate classical solutions is based on a Pohlmeyer type reduction\cite{Pohlmeyer,de Vega} in which the invariant dynamics is identified with the dynamics of sine-Gordon (or more generally of Toda) type. The other well known method is the dressing method, which was successfully applied for construction of magnon type solutions.\cite{SV}\cdash\cite{CDO}

In the study of AdS dynamics we follow the inverse scattering method where one generates string configurations starting with soliton solutions of the reduced field theory. This method was originally applied in Ref. \refcite{de Vega} for the study of string dynamics in de Sitter space. In Anti de Sitter space we have found a natural identification of soliton configurations, they were identified in Ref. \refcite{JJKV} with spikes of string foldings present in the Gubser-Klebanov-Polyakov-Kruczenski (GKP-K) solution. We will review this identification in what follows and give a more in depth discussion of it. The rotating string solutions of GKP-K simplify drastically in the limit $\omega=1$, when the rotational momentum is maximal. In that limit the spikes extend to the boundary of the AdS space but in the process some features of the solution get lost. We will describe at the sinh-Gordon level the intricacies of the $\omega=1$ limit and then proceed with the discussion of the $w>1$ case. While we concentrate on the two-soliton consideration we outline the construction for the (static) n-soliton case.

In the Euclidean worldsheet framework Alday and Maldacena have used minimal area classical string configurations to evaluate scattering amplitudes of Yang-Mills gluons\cite{AM} in terms of Wilson loops.\cite{KT} In this case the momenta of individual gluons specify a polygon geometry of the string worldsheet. It is then a challenging problem to construct minimal area string configurations with general n-polygon boundary conditions\cite{JKSV}\cdash\cite{Nastase}. This was accomplished in Ref. \refcite{AM} for the four-point case through an analytic continuation of Minkowski worldsheet solutions described above. One has the expectation that knowledge of more general Minkowski space solutions can therefore be of use for the gluon scattering problem also.

\section{AdS string as a $\sigma$-model}

We will concentrate in what follows on string dynamics in purely AdS spacetime. The string equations of motion in curved  spacetime can be formulated as generalized nonlinear $\sigma$ models provided one uses a conformal type gauge. Defining the $AdS_d$ space as $q^2=-q_{-1}^2-q_0^2+q_1^2+\cdots+q_{d-1}^2=-1$, the conformal gauge string equations are given by a noncompact $SO(d-1,2)$-symmetric $\sigma$ model with the action
\begin{equation}
A={\sqrt{\lambda} \over 2\pi}\int d\sigma d\tau \bigl(\partial q \cdot \partial q + \lambda (\sigma ,\tau)(q\cdot q+1)\bigr)
\end{equation}
where $\sigma,\tau$ are the Minkowski worldsheet coordinates, the equations of motion are
\begin{equation}
q_{\xi\eta}-(q_\xi \cdot q_\eta)q=0
\end{equation}
with $\xi=(\sigma+\tau)/2,\eta=(\sigma-\tau)/2$. In addition to guarantee the conformal gauge we have to impose the Virasoro conditions
\begin{equation}
q_\xi^2=q_\eta^2=0.
\end{equation}

It was demonstrated a number of years ago (by Pohlmeyer) that nonlinear sigma models subject to Virasoro type constraints can be reduced to known, integrable field equations of sine-Gordon (or Toda) type. This reduction is accomplished by concentrating on $SO(d-1,2)$ invariant sub-dynamics of the sigma model. The steps of the reduction were well described in Refs. \refcite{JJKV}--\refcite{de Vega} and consist in
the following. One starts by identifying first an appropriate set of basis vectors for the string coordinates. For $AdS_3$, the basis can be chosen as
\begin{equation}
e_i=(q,q_\xi,q_\eta,b)
\label{basis1}
\end{equation}
where $b$ is a fourth orthonormal vector, satisfying $b \cdot b=1,b \cdot q=b \cdot q_\xi=b \cdot q_\eta=0$. The reduced (invariant) scalar field is introduced through a scalar product $q_\xi \cdot q_\eta \equiv e^{\alpha(\xi,\eta)}$, and one proceeds to derive the equation of motion for $\alpha$, which reads
\begin{equation}
\alpha_{\xi\eta}-e^\alpha-uve^{-\alpha}=0
\label{sinh}
\end{equation}
where $u$ and $v$ are two additional (invariant) scalar fields given by $u=b \cdot q_{\xi\xi},v=b \cdot q_{\eta\eta}$. They are found to obey the equations $u_\eta=v_\xi=0$ and now the closed set of equations defines the generalized sinh-Gordon model.

One can next work out the equations obeyed by the elements of the basis, the derivatives of the vectors (\ref{basis1}) can be expressed in terms of the basis itself
\begin{equation}
{\partial e_i \over \partial \xi}=A_{ij}(\xi,\eta) e_j,~~{\partial e_i \over \partial \eta}=B_{ij}(\xi,\eta) e_j,
\end{equation}
where
\begin{equation}
A=\begin{pmatrix} 0 & 1 & 0 & 0 \cr 0 & \alpha_\xi & 0 & u \cr e^\alpha & 0 & 0 & 0 \cr 0 & 0 & -ue^{-\alpha} & 0 \end{pmatrix},~~B=\begin{pmatrix} 0 & 0 & 1 & 0 \cr e^\alpha & 0 & 0 & 0 \cr 0 & 0 & \alpha_\eta & v \cr 0 & -ve^{-\alpha} & 0 & 0 \end{pmatrix}.
\end{equation}
One finds therefore a linear system of differential equations for the vectors. The associated integrability condition reads
\begin{equation}
\partial_\eta A-\partial_\xi B+[A,B]=0.
\end{equation}
The integrability condition is then seen to generate the equations of motion corresponding to a  generalized sinh-Gordon theory. The vector equations on the other hand define the motion (and coordinates) of the string itself, they have to be solved, which leads to a scattering problem of Dirac type. This equations exhibit $SO(2,2)$ symmetry, and can be further simplified by redefining the orthonormal basis as\cite{Papanicolaou}
\begin{equation}
e_1=b,~~e_2={q_\xi+q_\eta \over \sqrt{2}e^{\alpha/2}},~~e_3={q_\xi-q_\eta \over \sqrt{2} i e^{\alpha/2}},~~e_4=iq.
\end{equation}
Then the $A,B$ matrices become
\begin{eqnarray}
A=\begin{pmatrix} 0 & -{u\over \sqrt{2}}e^{-\alpha/2} & {iu\over \sqrt{2}}e^{-\alpha/2} & 0 \cr {u\over \sqrt{2}}e^{-\alpha/2} & 0 & {i\over 2}\alpha_\xi & -{i\over \sqrt{2}}e^{\alpha/2} \cr -{iu\over \sqrt{2}}e^{-\alpha/2} & -{i\over 2}\alpha_\xi & 0 & {1\over \sqrt{2}}e^{\alpha/2} \cr 0 & {i\over \sqrt{2}}e^{\alpha/2} & -{1\over \sqrt{2}}e^{\alpha/2} & 0 \end{pmatrix}, \\
B=\begin{pmatrix} 0 & -{v\over \sqrt{2}}e^{-\alpha/2} & -{iv\over \sqrt{2}}e^{-\alpha/2} & 0 \cr {v\over \sqrt{2}}e^{-\alpha/2} & 0 & -{i\over 2}\alpha_\eta & -{i\over \sqrt{2}}e^{\alpha/2} \cr {iv\over \sqrt{2}}e^{-\alpha/2} & {i\over 2}\alpha_\eta & 0 & -{1\over \sqrt{2}}e^{\alpha/2} \cr 0 & {i\over \sqrt{2}}e^{\alpha/2}  & {1\over \sqrt{2}}e^{\alpha/2} & 0 \end{pmatrix}.
\end{eqnarray}
One now exploits the fact that $SO(2,2)=SO(2,1) \times SO(2,1)$, expanding the $A,B$ matrices in terms of two commuting sets of $SO(2,1)$ generators
\begin{equation}
A=w_{1,(+)}^i J_i + w_{1,(-)}^i K_i,~~~B=w_{2,(+)}^i J_i + w_{2,(-)}^i K_i,
\end{equation}
with $i=1,2,3$. Remember $SO(2,1)=SU(1,1)$, we can rewrite this problem in terms of the spinor representation of the $SU(1,1)$ group. Defining two spinors $\phi$ and $\psi$ satisfying the differential equations
\begin{eqnarray}
\phi_\xi&=&w_{1,(+)}^i\sigma_i\phi=A_1\phi,\hspace{.19in}\phi_\eta=w_{2,(+)}^i\sigma_i\phi=A_2\phi, \\
\psi_\xi&=&w_{1,(-)}^i\sigma_i\psi=B_1\psi,\hspace{.15in}\psi_\eta=w_{2,(-)}^i\sigma_i\psi=B_2\psi,
\end{eqnarray}
where $\sigma_i$ are the anti-Hermitian generators of $SU(1,1)$ group. The matrices $A_1,A_2,B_1,B_2$ can be found in Ref. \refcite{JJKV}. The  string solution is finally given by
\begin{eqnarray}
q_{-1} &=& {1\over 2}(\phi_1 \psi_1^* - \phi_2 \psi_2^*)+ c. c.~, \hspace{.3in} q_0={i\over 2}(\phi_1 \psi_1^* - \phi_2 \psi_2^*)+ c. c.~, \\
q_1 &=& {1\over 2} (\phi_2 \psi_1 - \phi_1 \psi_2) + c. c.~, \hspace{.32in} q_2={i\over 2} (\phi_2 \psi_1 - \phi_1 \psi_2) + c. c.~.
\end{eqnarray}

\section{GKP solution as a two-soliton configuration}

Gubser, Klebanov and Polyakov\cite{GKP} pointed out the relevance of semiclassical quantization in AdS giving the example of a large $S$ (spin angular momentum) rigidly rotating string. One constructs the GKP solution in the conformal gauge with the sigma model action
\begin{equation}
A={\sqrt{\lambda} \over 4\pi} \int d\tau d\sigma G_{ij} \partial_\alpha X^i \partial^\alpha X^j
\end{equation}
and the Virasoro constraints 
\begin{equation}
T_{++}=\partial_+ X^i \partial_+ X^j G_{ij}=0,~~~T_{--}=\partial_- X^i \partial_- X^j G_{ij}=0.
\end{equation}
The classical motion describing a rigid rotation of a folded closed string is given by the ansatz $t=c\tau,~\theta=c\omega\tau$ and  $\rho=\rho(\sigma)$. The Virasoro constraints give
\begin{equation}
\rho'^2=c^2(\cosh^2\rho-\omega^2\sinh^2\rho)
\label{GKPsolution}
\end{equation}
where the scaling constant $c$ is adjusted to define the period of $\sigma$. We can set $c=1$ and denote the position of the fold (spike) as $\sigma_0$. To demonstrate the stated correspondence with solitons we expand the solution (\ref{GKPsolution}) near the spike with $\omega=1+2\eta$, where $\eta \ll 1$, one finds
\begin{equation}
{\rho^\prime}^2 \sim e^{2\rho}(e^{-2\rho}-\eta).
\end{equation}
Denoting $u=e^{-\rho}$, we have ${u^\prime}^2 \sim u^2-\eta$. Consider the boundary condition $u_0=e^{-\rho_0}$ at $\sigma=\sigma_0$, one finds
\begin{equation}
\rho(\sigma)=-\ln \bigl(\sqrt{\eta} \cosh (\sigma-\sigma_0) \bigr),
\end{equation}
so that
\begin{equation}
\alpha \equiv \ln(q_\xi \cdot q_\eta)=\ln(2\rho'^2)=\ln(2\tanh^2\sigma).
\end{equation}
This is exactly the one-soliton solution to the sinh-Gordon equation $\alpha_{\xi\eta}-e^\alpha+4 e^{-\alpha}=0$.\cite{JJKV} The closed GKP solution has two folds (spikes) and therefore corresponds in the sinh-Gordon picture to a two-soliton configuration. We will next describe its construction starting from the solutions of the sinh-Gordon system.

\section{AdS string solutions}

\begin{figure}[h]
\centerline{\psfig{file=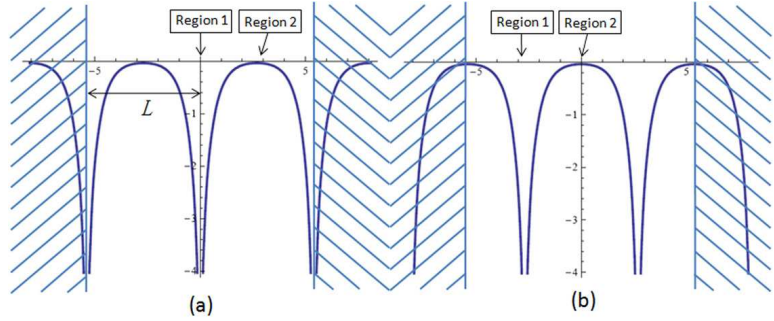,width=0.8\textwidth}}
\caption{(a) First periodic sinh-Gordon solution $\hat{\alpha}_1$ when $k=0.964$; (b) Second periodic sinh-Gordon solution $\hat{\alpha}_2$ when $k=0.964$. They are related by a translation of $\sigma \rightarrow \sigma + \sqrt{k}K(k)$.}
\label{f1}
\end{figure}

Consider the generalized sinh-Gordon equation (\ref{sinh}), making a shift of the field $\alpha=\hat{\alpha}+{1 \over 2}\ln(-uv)$, we have
\begin{equation}
\hat{\alpha}_{\xi \eta}-2\sqrt{-uv}\sinh\hat{\alpha}=0.
\end{equation}
In the case of $u=2,v=-2$, we consider the periodic solution
\begin{equation}
\hat{\alpha}_1=\ln[k~{\rm sn}^2({\sigma \over \sqrt{k}}, k)]
\end{equation}
with periodicity $L=2\sqrt{k}K(k)$ where $K(k)$ is the elliptic function and $k$ is a parameter with $0<k<1$ (see Fig. \ref{f1} (a)). The spinors are found to be
\begin{eqnarray}
\phi_1&=&{1 \over 2}\exp\Bigl[-{i\over2}{1+k \over \sqrt{k}} \tau\Bigr]\Bigl( \sqrt{{1+k{\rm sn}^2+{\rm cn}~{\rm dn} \over (1+k) {\rm sn}}}+\sqrt{{(1+k) {\rm sn} \over 1+k{\rm sn}^2+{\rm cn}~{\rm dn}}}\Bigr) \\
\phi_2&=&{1 \over 2}\exp\Bigl[-{i\over2}{1+k \over \sqrt{k}} \tau\Bigr]\Bigl( \sqrt{{1+k{\rm sn}^2+{\rm cn}~{\rm dn} \over (1+k) {\rm sn}}}-\sqrt{{(1+k) {\rm sn} \over 1+k{\rm sn}^2+{\rm cn}~{\rm dn}}}\Bigr) \\
\psi_1&=&{1 \over 2}\exp\Bigl[-{i\over2}{1-k \over \sqrt{k}} \tau\Bigr]\Bigl( \sqrt{{1-k{\rm sn}^2+{\rm cn}~{\rm dn} \over (1-k) {\rm sn}}}+\sqrt{{(1-k) {\rm sn} \over 1-k{\rm sn}^2+{\rm cn}~{\rm dn}}}\Bigr)
\end{eqnarray}
\begin{eqnarray}
\psi_2&=&{1 \over 2}\exp\Bigl[-{i\over2}{1-k \over \sqrt{k}} \tau\Bigr]\Bigl( \sqrt{{1-k{\rm sn}^2+{\rm cn}~{\rm dn} \over (1-k) {\rm sn}}}-\sqrt{{(1-k) {\rm sn} \over 1-k{\rm sn}^2+{\rm cn}~{\rm dn}}}\Bigr)
\end{eqnarray}
where cn, sn and dn are Jacobi elliptic functions with respect to $({\sigma \over \sqrt{k}}, k)$. The string solution is given by
\begin{equation}
q_1=\begin{pmatrix} {1 \over \sqrt{1-k^2}}{\rm dn}({\sigma \over \sqrt{k}},k) \cos \sqrt{k}\tau \cr {1 \over \sqrt{1-k^2}}{\rm dn}({\sigma \over \sqrt{k}},k) \sin \sqrt{k}\tau \cr {k \over \sqrt{1-k^2}}{\rm cn}({\sigma \over \sqrt{k}},k) \cos {1\over \sqrt{k}}\tau \cr {k \over \sqrt{1-k^2}}{\rm cn}({\sigma \over \sqrt{k}},k) \sin {1\over \sqrt{k}}\tau \end{pmatrix}.
\label{solution1}
\end{equation}
For $0<k<1$, this solution is well defined and periodic so that the string is closed. The above solution is divergent at $k=1$. This does not mean there is no regular solution when $k=1$. In this case, the sinh-Gordon solution becomes
\begin{equation}
\hat{\alpha}_{k=1}=\ln[\tanh^2 \sigma],
\end{equation}
we can choose different normalization coefficients of the $\psi$ spinor and the string solution is given by\cite{JJKV}
\begin{equation}
q_{k=1}={1 \over 2\sqrt{2} \cosh \sigma} \begin{pmatrix} 2\tau\cos\tau-\sin\tau(\cosh 2\sigma+2) \cr 2\tau\sin\tau+\cos\tau(\cosh 2\sigma+2)\cr -2\tau\cos\tau+\sin\tau\cosh 2\sigma \cr -2\tau\sin\tau-\cos\tau\cosh 2\sigma \end{pmatrix}.
\end{equation}
In the limit $k=1$ the string parameter space is decompactified, we see an infinite string solution which touches the boundary of AdS. Due to the nonvanishing boundary condition at infinity, there is momentum flow at the boundary of the string and the energy is not conserved. These were the features of the one soliton string configuration given in Ref. \refcite{JJKV}.

It is also interesting to note another limit which leads to vacuum at $k=1$. First, we shift the sinh-Gordon solution by half the period $\sigma \rightarrow \sigma+\sqrt{k}K(k)$ and obtain
\begin{equation}
\hat{\alpha}_2=\ln[k~{\rm cn}^2({\sigma \over \sqrt{k}}, k)~{\rm nd}^2({\sigma \over \sqrt{k}}, k)]
\end{equation}
(see Fig. \ref{f1}(b)). We note that this solution reduces to vacuum $\hat{\alpha}=0$ in the limit of $k=1$. The corresponding string solution is given by
\begin{equation}
q_2=\begin{pmatrix} {\rm nd}({\sigma \over \sqrt{k}},k) \cos \sqrt{k}\tau \cr {\rm nd}({\sigma \over \sqrt{k}},k) \sin \sqrt{k}\tau \cr k~{\rm sd}({\sigma \over \sqrt{k}},k) \cos {1 \over \sqrt{k}} \tau \cr k~{\rm sd}({\sigma \over \sqrt{k}},k) \sin {1 \over \sqrt{k}} \tau \end{pmatrix}.
\label{solution2}
\end{equation}
As we expected, in the limit of $k=1$, this solution reduces to the vacuum string solution in Ref. \refcite{JJKV}. These two limits at $k=1$ are essentially expanding the sinh-Gordon solution around the soliton (region 1) and the vacuum (region 2), respectively, as shown in Fig. \ref{f1}.

Do the rescaling $\sqrt{k}\tau \rightarrow \tau, \sqrt{k}\sigma \rightarrow \sigma$ and write $k=1/\omega$, the string solutions (\ref{solution1}) and (\ref{solution2}) correspond to the minus or plus solution of (\ref{GKPsolution}). Remember that the GKP solution is a two-soliton configuration with the period of $\sigma \in [0,2L]$ where $L={2 \over \omega}K({1 \over \omega})$ after rescaling. The energy and angular momentum are exactly calculated to be
\begin{eqnarray}
E&=&{\sqrt{\lambda} \over 2\pi}\int_0^{2L}d\sigma \cosh^2 \rho={2\sqrt{\lambda} \over \pi}\Bigl[{\omega \over \omega^2-1}E\Bigl({1 \over \omega}\Bigr) \Bigr], \\
S&=&{\sqrt{\lambda} \over 2\pi}\int_0^{2L}d\sigma~\omega \sinh^2 \rho={2\sqrt{\lambda} \over \pi}\Bigl[{\omega^2 \over \omega^2-1}E\Bigl({1 \over \omega}\Bigr)-K\Bigl({1 \over \omega}\Bigr)\Bigr].
\end{eqnarray}
where ${\rm E}({1 \over \omega})$ and ${\rm K}({1 \over \omega})$ are elliptic functions. Therefore,
\begin{equation}
E-\omega S={2\omega \sqrt{\lambda} \over \pi}\Bigl[K\Bigl({1 \over \omega}\Bigr)-E\Bigl({1 \over \omega}\Bigr)\Bigr].
\end{equation}
For long strings, $\omega = 1+2\eta$ where $\eta \ll 1$, we can expand the elliptic functions and get
\begin{equation}
E-S={\sqrt{\lambda}\over \pi} \ln S+\cdots
\end{equation}
which agrees with Ref. \refcite{GKP}.

\section{N-soliton construction}

\begin{figure}[t]
\centerline{\psfig{file=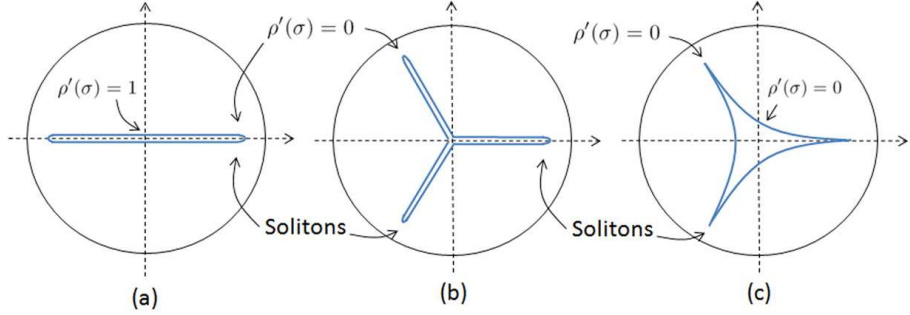,width=0.95\textwidth}}
\caption{(a) GKP two-soliton configuration plotted in the plane $x=\rho\cos\theta,y=\rho\sin\theta$ where $\rho,\theta$ are the global coordinates; (b) A attempt to construct the GKP type three-soliton solution; (c) Kruczenski's three-spike string solution.}
\label{f2}
\end{figure}

\begin{figure}[h]
\centerline{\psfig{file=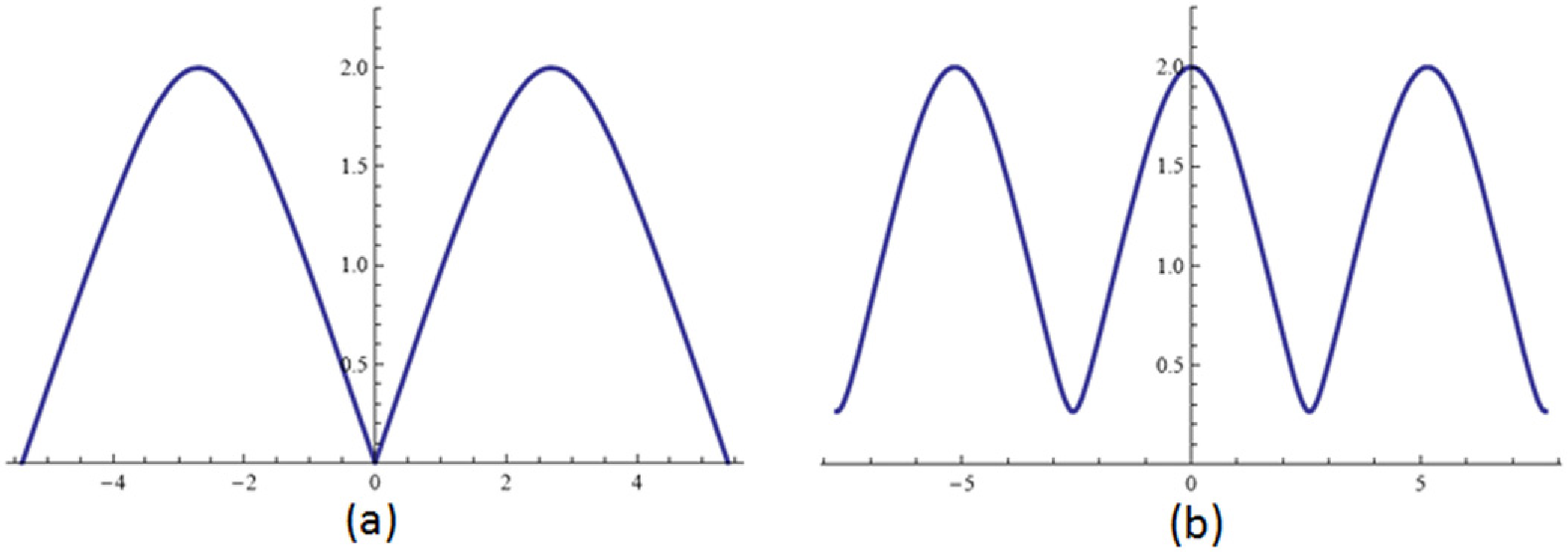,width=0.8\textwidth}}
\caption{(a) GKP $\rho$ as a function of $\sigma$ when $k=0.964$; (b) Kruczenski $\rho$ as a function of $\sigma$ when $\rho_1=2,\rho_0=0.2688735$.}
\label{f3}
\end{figure}

Following the discussion of the periodic two-soliton (GKP) solution (see Fig. \ref{f2}(a)), we would like to consider a possible n-soliton generalization. In terms of a naive gluing procedure one would be lead to a configuration given  in Fig. \ref{f2}(b). At the sinh-Gordon level the analogue construction is simple, one essentially extends the length of the space to accommodate n static solitons.  However it does not follow that this configuration is continuous and nonsingular in the string configuration space. We have the fact that at the center of AdS space where $\rho=0$, there results only two solutions of (\ref{GKPsolution}): $\rho'(\sigma)=\pm 1$ (see Fig. \ref{f3}(a)). This allows for two angles: 0 and $\pi$ which corresponds to the gluing in the case of a GKP configuration. This implies that the string extends along a straight line. In the physical gauge $t=\tau,\theta=\omega \tau+\sigma$ Kruczenski\cite{Kruczenski} succeeded with the construction of a n-spike static configurations (see Fig. \ref{f2}(c)). This indeed regulates the naive configuration of Fig. \ref{f2}(b). As we will discuss this solution can be obtained by lifting the minimum value of $\rho$ to be $\rho_0$ where $\rho'(\sigma)=0$ (see Fig. \ref{f3}(b)) and gluing n spikes at that point. It will correspond to sinh-Gordon solutions with non-zero boundary conditions.

We can reproduce Kruczenski's solution in the conformal gauge by making the ansatz $t=\tau+f(\sigma),\theta=\omega \tau+g(\sigma),\rho=\rho(\sigma)$. The equations of motion and Virasoro constraints can be solved by
\begin{eqnarray}
f'(\sigma)&=&{\omega \sinh 2\rho_0 \over 2\cosh^2 \rho},~~~~~g'(\sigma)={\sinh 2\rho_0 \over 2\sinh^2 \rho}, \cr
\rho'^2(\sigma)&=&{(\cosh^2\rho-\omega^2\sinh^2\rho)(\sinh^2 2\rho-\sinh^2 2\rho_0) \over \sinh^2 2\rho}.
\label{Krucsolution}
\end{eqnarray}
Near the spike, we have $\rho \sim \rho_1 \equiv {\rm arccoth}\omega$, further assume $\rho_1 \gg \rho_0$, we can recover (\ref{GKPsolution}) from (\ref{Krucsolution}). Therefore, a soliton is located at each spike and the finite n-spike string solution is a n-soliton configuration. The differential equation (\ref{Krucsolution}) can be solved to be
\begin{equation}
\rho={1\over2}\rm{arccosh}\bigl(\cosh 2\rho_1 {\rm cn}^2(u,k) + \cosh 2\rho_0 {\rm sn}^2 (u,k)\bigr)
\label{first}
\end{equation}
where
\begin{equation}
u \equiv \sqrt{\cosh 2\rho_1+\cosh 2\rho_0 \over \cosh 2\rho_1-1} \sigma,~~~k \equiv \sqrt{\cosh 2\rho_1 - \cosh 2\rho_0 \over \cosh 2\rho_1 + \cosh 2\rho_0},
\end{equation}
and $\rho_0,\rho_1$ are the minimum and maximum values of $\rho$, respectively. The gauge transformation functions are found to be
\begin{eqnarray}
f&=&{\sqrt{2} \omega \sinh 2\rho_0 \sinh \rho_1 \over (\cosh 2\rho_1+1) \sqrt{\cosh 2\rho_1+\cosh 2\rho_0}}\Pi\Bigl({\cosh 2\rho_1-\cosh 2\rho_0 \over \cosh 2\rho_1 +1},x,k\Bigr) \\
g&=&{\sqrt{2} \sinh 2\rho_0 \sinh \rho_1 \over (\cosh 2\rho_1-1) \sqrt{\cosh 2\rho_1+\cosh 2\rho_0}}\Pi\Bigl({\cosh 2\rho_1-\cosh 2\rho_0 \over \cosh 2\rho_1 -1},x,k\Bigr)
\end{eqnarray}
where $x={\rm am}(u,k)$ and $\Pi(n,x,k)$ is the incomplete elliptic integral.

The sinh-Gordon solution corresponding to the n-spike solution is
\begin{equation}
\alpha=\ln 2(\cosh^2 \rho-\omega^2 \sinh^2 \rho)
\end{equation}
with $uv=(1-\omega^2)^2\sinh^2 2\rho_0-4\omega^2$. After the change of variables, we get
\begin{equation}
\hat{\alpha}=\ln \bigl[k~{\rm sn}^2(u,k)\bigr]
\end{equation}
which, at the limit of $\rho_0=0$, reduces to $\hat{\alpha}_1$ (after scaling). There is also a shifted solution corresponding to $\hat{\alpha}_2$. In this sense, we say the Kruczenski's solution can be obtained by lifting the minimum value of $\rho$ as compared to the GKP solution. A parallel statement holds for the associated sinh-Gordon soliton solutions.

For completeness we give the energy and angular momentum of these configurations. They can be exactly computed to read
\begin{eqnarray}
E&=&{n\sqrt{\lambda} \over \pi}\sqrt{{\rm ch} 2\rho_1 - 1 \over {\rm ch} 2\rho_1+{\rm ch} 2\rho_0} \Bigl[{1 \over 2}({\rm ch} 2\rho_1+{\rm ch} 2\rho_0){\rm E}(k)-{\rm sh}^2 \rho_0 {\rm K}(k)\Bigr], \\
S&=&{n\omega\sqrt{\lambda} \over \pi}\sqrt{{\rm ch} 2\rho_1 - 1 \over {\rm ch} 2\rho_1+{\rm ch} 2\rho_0} \Bigl[{1 \over 2}({\rm ch} 2\rho_1+{\rm ch} 2\rho_0){\rm E}(k)-{\rm ch}^2 \rho_0 {\rm K}(k)\Bigr], \\
E&-&\omega S={n \sqrt{\lambda} \over \pi} \sqrt{{\rm ch} 2\rho_1+{\rm ch} 2\rho_0 \over {\rm ch} 2\rho_1 - 1}\Bigl[K(k)-E(k)\Bigr].
\end{eqnarray}
In the limit of $\omega \rightarrow 1$ and assume $\rho_1 \gg \rho_0$, we find
\begin{equation}
E-S=n {\sqrt{\lambda} \over 2\pi} \ln S +\cdots
\end{equation}
showing agreement with the result of Ref. \refcite{Kruczenski}.

\section{Conclusion}
In this contribution we reviewed and generalized some simple classical solutions for strings moving in AdS spacetime. We have studied in depth on the so-called spiky string configurations\cite{GKP,Kruczenski} and their properties. We reviewed the approach of Ref. \refcite{JJKV} which is based on the identification of string spikes with soliton configurations. This explains the usefulness of the (inverse) scattering technique in constructing string configurations of this type. In the review we paid particular attention to the distinction between compact and non-compact string parameter space solutions elaborating on the limit relating the two. The soliton and (inverse) scattering techniques are expected to be of definite use for studying more general sets of solutions (and their dynamical properties). It is also possible that they will be of use for addressing the very interesting and highly nontrivial `platoux' problem of the Euclidean worldsheet string theory.

\section*{Acknowledgments}

We are grateful to C. Kalousios and A. Volovich for collaboration on which this work is based. We would also like to thank M. Abbott, I. Aniceto and M. Spradlin for comments and discussions. One of us (AJ) is grateful to Hiroshi Itoyama, Hikaru Kawai and Masao Ninomiya for their kind hospitality. This work is supported by the Department of Energy under contract DE-FG02-91ER40688.


\end{document}